\pdfoutput=1
\documentclass[12pt,a4paper]{article}

\usepackage{lineno}  
\usepackage{graphicx}  
\usepackage{xspace} 
\usepackage{color}
\usepackage{colortbl}
\usepackage{amsmath} 
\usepackage{ifthen} 

\usepackage{cite}
\usepackage{placeins}
\usepackage{subfigure}

\newboolean{pdflatex}
\setboolean{pdflatex}{true} 
\newboolean{articletitles}
\setboolean{articletitles}{true} 

\newboolean{uprightparticles}
\setboolean{uprightparticles}{false} 

\usepackage{amssymb}
\usepackage{amsfonts}
\usepackage{upgreek} 

\usepackage{hyperref}    
\usepackage[all]{hypcap} 

\def\lhcb {LHCb\xspace}
\def\ux85 {UX85\xspace}

\ifthenelse{\boolean{uprightparticles}}%
{

 \def\Ppi         {\ensuremath{\uppi}\xspace}

 \def\PDelta      {\ensuremath{\Delta}\xspace}                 
 \def\PXi      {\ensuremath{\Xi}\xspace}                 
 \def\PLambda      {\ensuremath{\Lambda}\xspace}                 
 \def\PSigma      {\ensuremath{\Sigma}\xspace}                 
 \def\POmega      {\ensuremath{\Omega}\xspace}                 
 \def\PUpsilon      {\ensuremath{\Upsilon}\xspace}                 
 
 \def\PB      {\ensuremath{\mathrm{B}}\xspace}                 
                  
 \def\PD      {\ensuremath{\mathrm{D}}\xspace}

 \def\PK      {\ensuremath{\mathrm{K}}\xspace}

 \def\Pi      {\ensuremath{\mathrm{i}}\xspace}

 \def\Ps      {\ensuremath{\mathrm{s}}\xspace}

}
{

 \def\Ppi         {\ensuremath{\pi}\xspace}

 \mathchardef\PDelta="7101
 \mathchardef\PXi="7104
 \mathchardef\PLambda="7103
 \mathchardef\PSigma="7106
 \mathchardef\POmega="710A
 \mathchardef\PUpsilon="7107
                  
 \def\PB      {\ensuremath{B}\xspace}                 
                  
 \def\PD      {\ensuremath{D}\xspace}

 \def\PK      {\ensuremath{K}\xspace}

 \def\Pi      {\ensuremath{i}\xspace}

 \def\Ps      {\ensuremath{s}\xspace}

}

\def\squark    {\ensuremath{\Ps}\xspace}

\def\pion  {\ensuremath{\Ppi}\xspace}

\def\pim   {\ensuremath{\pion^-}\xspace}

\def\kaon  {\ensuremath{\PK}\xspace}
  \def\Kbar  {\kern 0.2em\overline{\kern -0.2em \PK}{}\xspace}

\def\Kz    {\ensuremath{\kaon^0}\xspace}
\def\Kzb   {\ensuremath{\Kbar^0}\xspace}
\def\KzKzb {\ensuremath{\Kz \kern -0.16em \Kzb}\xspace}
\def\Kp    {\ensuremath{\kaon^+}\xspace}
\def\Km    {\ensuremath{\kaon^-}\xspace}

\def\KpKm  {\ensuremath{\Kp \kern -0.16em \Km}\xspace}

  \def\Dbar    {\kern 0.2em\overline{\kern -0.2em \PD}{}\xspace}
\def\D       {\ensuremath{\PD}\xspace}

\def\Dz      {\ensuremath{\D^0}\xspace}
\def\Dzb     {\ensuremath{\Dbar^0}\xspace}
\def\DzDzb   {\ensuremath{\Dz {\kern -0.16em \Dzb}}\xspace}
\def\Dp      {\ensuremath{\D^+}\xspace}
\def\Dm      {\ensuremath{\D^-}\xspace}

\def\DpDm    {\ensuremath{\Dp {\kern -0.16em \Dm}}\xspace}

\def\B       {\ensuremath{\PB}\xspace}
  \def\Bbar    {\kern 0.18em\overline{\kern -0.18em \PB}{}\xspace}

\def\Bz      {\ensuremath{\B^0}\xspace}

\def\Bd      {\ensuremath{\B^0}\xspace}
\def\Bs      {\ensuremath{\B^0_\squark}\xspace}
\def\Bsb     {\ensuremath{\Bbar^0_\squark}\xspace}

  \def\Y#1S{\ensuremath{\PUpsilon{(#1S)}}\xspace}

\newcommand{\decay}[2]{\ensuremath{#1\!\to #2}\xspace}         

\def\to                 {\ensuremath{\rightarrow}\xspace}

\newcommand{\tauBd}{\ensuremath{\tau_{\Bd}}\xspace}

\def\CP                {\ensuremath{C\!P}\xspace}

\newcommand{\DGs}{\ensuremath{\Delta\Gamma_{\squark}}\xspace}

\newcommand{\Gs}{\ensuremath{\Gamma_{\squark}}\xspace}

\newcommand{\GL}{\ensuremath{\Gamma_{\rm L}}\xspace}
\newcommand{\GH}{\ensuremath{\Gamma_{\rm H}}\xspace}

\def\BdToKpi      {\decay{\Bd}{\Kp\pim}}
\def\BsToKK       {\decay{\Bs}{\Kp\Km}}

\def\AT#1     {\ensuremath{A_{\mathrm{T}}^{#1}}\xspace}           

\def\C#1      {\ensuremath{\mathcal{C}_{#1}}\xspace}                       
\def\Cp#1     {\ensuremath{\mathcal{C}_{#1}^{'}}\xspace}                    
\def\Ceff#1   {\ensuremath{\mathcal{C}_{#1}^{\mathrm{(eff)}}}\xspace}        
\def\Cpeff#1  {\ensuremath{\mathcal{C}_{#1}^{'\mathrm{(eff)}}}\xspace}       
\def\Ope#1    {\ensuremath{\mathcal{O}_{#1}}\xspace}                       
\def\Opep#1   {\ensuremath{\mathcal{O}_{#1}^{'}}\xspace}                    

\newcommand{\tev}{\ensuremath{\mathrm{\,Te\kern -0.1em V}}\xspace}
\newcommand{\gev}{\ensuremath{\mathrm{\,Ge\kern -0.1em V}}\xspace}
\newcommand{\mev}{\ensuremath{\mathrm{\,Me\kern -0.1em V}}\xspace}
\newcommand{\kev}{\ensuremath{\mathrm{\,ke\kern -0.1em V}}\xspace}
\newcommand{\ev}{\ensuremath{\mathrm{\,e\kern -0.1em V}}\xspace}
\newcommand{\gevc}{\ensuremath{{\mathrm{\,Ge\kern -0.1em V\!/}c}}\xspace}
\newcommand{\mevc}{\ensuremath{{\mathrm{\,Me\kern -0.1em V\!/}c}}\xspace}
\newcommand{\gevcc}{\ensuremath{{\mathrm{\,Ge\kern -0.1em V\!/}c^2}}\xspace}
\newcommand{\gevgevcccc}{\ensuremath{{\mathrm{\,Ge\kern -0.1em V^2\!/}c^4}}\xspace}
\newcommand{\mevcc}{\ensuremath{{\mathrm{\,Me\kern -0.1em V\!/}c^2}}\xspace}

\def\invfb   {\ensuremath{\mbox{\,fb}^{-1}}\xspace}

\def\ps   {\ensuremath{{\rm \,ps}}\xspace}

\def\gsim{{~\raise.15em\hbox{$>$}\kern-.85em
          \lower.35em\hbox{$\sim$}~}\xspace}
\def\lsim{{~\raise.15em\hbox{$<$}\kern-.85em
          \lower.35em\hbox{$\sim$}~}\xspace}

\def\evtgen     {\mbox{\textsc{EvtGen}}\xspace}
\def\pythia     {\mbox{\textsc{Pythia}}\xspace}

\def\geant      {\mbox{\textsc{Geant4}}\xspace}

\def\photos     {\mbox{\textsc{Photos}}\xspace}

\def\tell1  {TELL1\xspace}
\def\ukl1   {UKL1\xspace}

\newcommand{\ADGs}      {\ensuremath{{\cal A}_{\Delta\Gamma_s}}\xspace}
\newcommand{\tauBsToKK} {\ensuremath{\tau_{KK }}\xspace}

\usepackage{mciteplus}

\begin{document}

\renewcommand{\thefootnote}{\fnsymbol{footnote}}
\setcounter{footnote}{1}


\begin{titlepage}
\pagenumbering{roman}

\vspace*{-1.5cm}
\centerline{\large EUROPEAN ORGANIZATION FOR NUCLEAR RESEARCH (CERN)}
\vspace*{1.5cm}
\hspace*{-0.5cm}
\begin{tabular*}{\linewidth}{lc@{\extracolsep{\fill}}r}
\ifthenelse{\boolean{pdflatex}}
{\vspace*{-2.7cm}\mbox{\!\!\!\includegraphics[width=.14\textwidth]{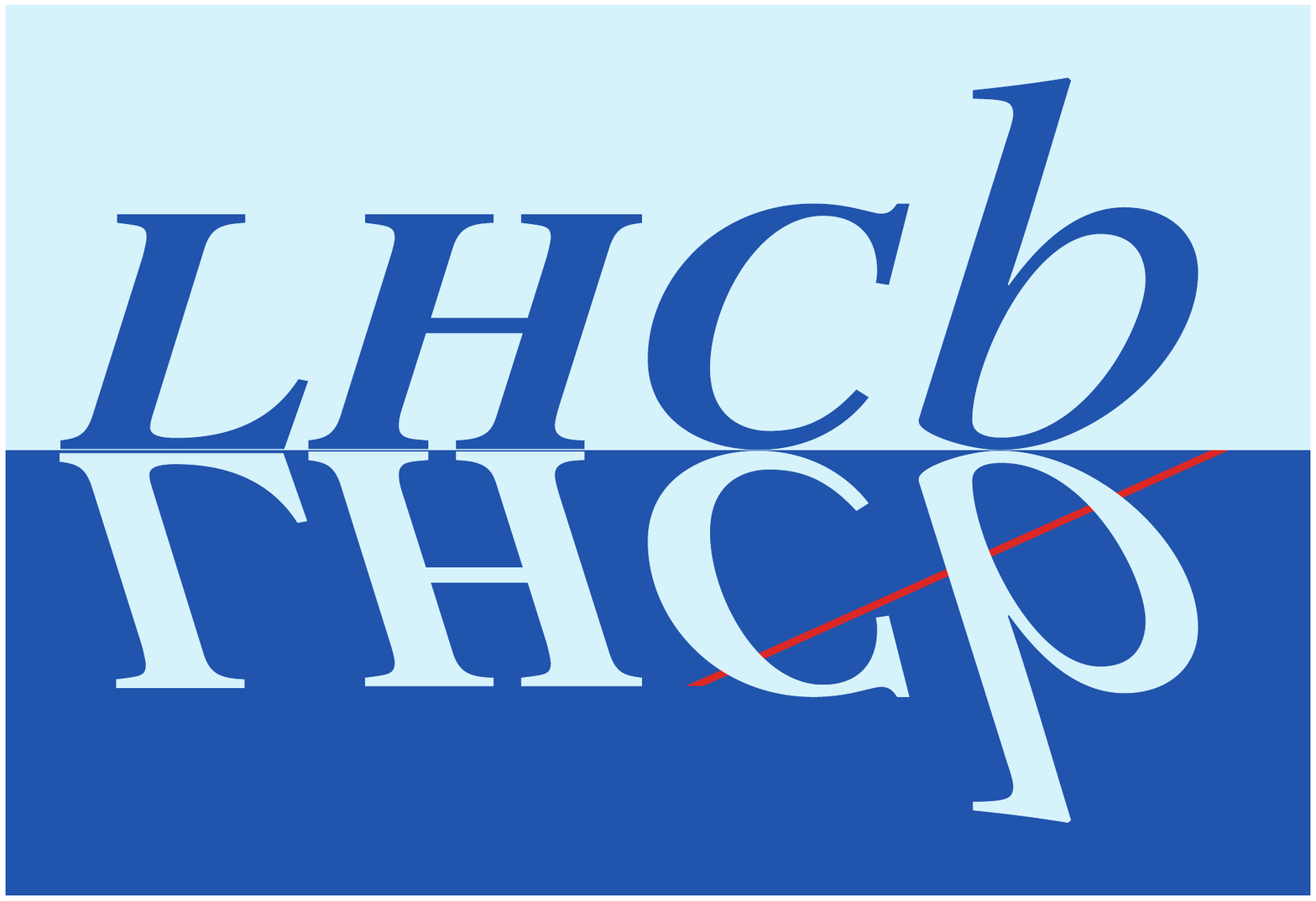}} & &}%
{\vspace*{-1.2cm}\mbox{\!\!\!\includegraphics[width=.12\textwidth]{lhcb-logo.eps}} & &}%
\\
 & & CERN-PH-EP-2012-172 \\  
 & & LHCb-PAPER-2012-013 \\  
 & & 19 September 2012 \\ 
 & & \\
\end{tabular*}

\vspace*{4.0cm}

{\bf\boldmath\huge
\begin{center}
  Measurement of the effective \BsToKK lifetime
\end{center}
}

\vspace*{2.0cm}

\begin{center}
The LHCb collaboration\footnote{Authors are listed on the following pages.}
\end{center}

\vspace{\fill}

\begin{abstract}
  \noindent
A precise determination of the effective \BsToKK lifetime can be used to constrain contributions from physics beyond the Standard Model in the \Bs meson system.
Conventional approaches select \B meson decay products that are significantly displaced from the \B meson production vertex. As a consequence, \B mesons with low decay times are suppressed, introducing a bias to the decay time spectrum which must be corrected.
This analysis uses a technique that explicitly avoids a lifetime bias by using a neural network based trigger and event selection. Using 1.0~\invfb of data recorded by the LHCb experiment, the effective \BsToKK lifetime is measured as 
$1.455 \pm 0.046 \; \mathrm{(stat.)} \pm  0.006 \; \mathrm{(syst.)}  \, \mathrm{ps}.$
\end{abstract}

\vspace*{2.0cm}

\begin{center}
  Submitted to Physics Letters B.
\end{center}

\vspace{\fill}

\end{titlepage}


\newpage
\setcounter{page}{2}
\mbox{~}
\newpage

\centerline{\large\bf LHCb collaboration}
\begin{flushleft}
\small
R.~Aaij$^{38}$, 
C.~Abellan~Beteta$^{33,n}$, 
A.~Adametz$^{11}$, 
B.~Adeva$^{34}$, 
M.~Adinolfi$^{43}$, 
C.~Adrover$^{6}$, 
A.~Affolder$^{49}$, 
Z.~Ajaltouni$^{5}$, 
J.~Albrecht$^{35}$, 
F.~Alessio$^{35}$
% ...

{\footnotesize \it
% ...
$ ^{5}$Clermont Universit\'{e}, Universit\'{e} Blaise Pascal, CNRS/IN2P3, LPC, Clermont-Ferrand, France\\
$ ^{6}$CPPM, Aix-Marseille Universit\'{e}, CNRS/IN2P3, Marseille, France\\
% ...
$ ^{11}$Physikalisches Institut, Ruprecht-Karls-Universit\"{a}t Heidelberg, Heidelberg, Germany\\
% ...
$ ^{33}$Universitat de Barcelona, Barcelona, Spain\\
$ ^{34}$Universidad de Santiago de Compostela, Santiago de Compostela, Spain\\
$ ^{35}$European Organization for Nuclear Research (CERN), Geneva, Switzerland\\
% ...
$ ^{38}$Nikhef National Institute for Subatomic Physics, Amsterdam, The Netherlands\\
% ...
$ ^{43}$H.H. Wills Physics Laboratory, University of Bristol, Bristol, United Kingdom\\
% ...
$ ^{49}$Oliver Lodge Laboratory, University of Liverpool, Liverpool, United Kingdom\\
% ...
$ ^{n}$LIFAELS, La Salle, Universitat Ramon Llull, Barcelona, Spain\\
% ...
}
\end{flushleft}

\cleardoublepage


\renewcommand{\thefootnote}{\arabic{footnote}}
\setcounter{footnote}{0}



\pagestyle{plain} 
\setcounter{page}{1}
\pagenumbering{arabic}


%

\section{Introduction}
\label{sec:intro}
The study of charmless $b$-hadron decays can be used to explore the phase structure of the CKM matrix and to search for indirect evidence of physics beyond the Standard Model (SM). A measurement of the effective lifetime of the \BsToKK decay (charge conjugate modes are implied throughout) is of considerable interest as it is sensitive to new physical phenomena affecting the \Bs mixing phase and entering the decay at loop level~\cite{bib:Grossman,Lenz:2006hd,bib:Fleischer1,bib:Fleischer2}.

The \BsToKK decay was first observed by the CDF collaboration~\cite{bib:CDFBs} and the most precise 
measurement to date of the effective lifetime was made by the \lhcb collaboration using data taken during 2010 \cite{TauBhhLHCbPLB}.
A detailed theoretical description of the \BsToKK decay can
be found in Refs.~\cite{bib:Fleischer1,bib:Fleischer2}.  When the initial flavour of the \Bs meson is unknown the decay time distribution can be written as
\begin{eqnarray}
  \label{eqn:dblExp}
  \Gamma(t) & \propto & \left ( 1 - \ADGs  \right )e^{-\Gamma_L t} + \left ( 1 + \ADGs \right )e^{-\Gamma_H t}.
\end{eqnarray}
The quantities \GH and \GL are the decay widths of the heavy and light \Bs mass eigenstates and $\DGs = \GL - \GH$ is the decay width difference. The parameter \ADGs is defined as $\ADGs = -2 {\rm Re}(\lambda)/\left(1 + |\lambda|^2\right)$ where $\lambda = (q/p)(\bar{A}/A)$, where the complex coefficients $p$ and $q$ define
the mass eigenstates of the \Bs--\Bsb system in terms of the flavour
eigenstates (see {\it e.g.}, Ref.~\cite{Nakamura:2010zzi}) and $A$
($\bar{A}$) is the amplitude for a \Bs (\Bsb) meson to decay to the
$\Kp\Km$ final state.

If the decay time distribution given by Eq.~\ref{eqn:dblExp} is fitted with a single exponential function the \textit{effective lifetime} is given by~\cite{bib:Hartkorn99}
\begin{equation}
  \label{eq:tauKKPred}
  \tauBsToKK =  \frac{\tau_{\Bs}}{1-y_s^2} \left [\frac{1 +
      2\ADGs y_s + y_s^2}{1 + \ADGs y_s} \right]
 = \tau_{\Bs} \left( 1 + \ADGs y_s + \mathcal{O}(y_s^2) \right),
\end{equation}
where $\tau_{\Bs} = 2 / \left ( \Gamma_H + \Gamma_L \right ) = \Gs^{-1}$ and $y_s = \DGs/2\Gs$.
The $\Kp\Km$ final state is \CP-even and so in the SM the decay is dominated by the light mass eigenstate such that $\ADGs = -0.972 \pm 0.012$ \cite{bib:Fleischer1,deBruyn:2012wj} and the effective lifetime thus is approximately equal to $\GL^{-1}$. Adopting the approach of Ref.~\cite{bib:Fleischer1} and using the world averages of $\Gamma_s$ and $\Delta\Gamma_s$~\cite{Asner:2010qj} and the SM prediction of \ADGs, the effective lifetime is predicted to be $\tauBsToKK = 1.40 \pm 0.02 \ps$. However, the \BsToKK decay is dominated by penguin diagrams and so is sensitive to physics beyond the SM entering at loop level, which may affect \ADGs. The measurement is also sensitive to new physics contributions to the \Bs mixing phase which in turn affects \DGs \cite{bib:Fleischer3}. Deviations from this prediction will therefore provide evidence of new physics.

Conventional selections exploit the long lifetimes of $b$-hadrons by requiring that their decay products are significantly displaced from the primary
interaction point. However, this  introduces a time-dependent acceptance of the selected $b$-hadron candidates which needs to be taken into account in the analysis. This paper describes a technique based on neural networks which avoids such acceptance effects. Only properties independent of the decay time are used to discriminate between signal and background. To exploit the available information, including the correlations between variables, several neural networks are used in a dedicated trigger and event selection.

\section{The LHCb experiment and simulation software}
\label{sec:Detector}
The \BsToKK lifetime is measured using $1.0~\invfb$ of \textit{pp} collision data collected by the \lhcb detector at a centre of mass energy of $\sqrt{s} = 7$ TeV during 2011. The \lhcb detector \cite{Alves:2008zz} is a single arm spectrometer with a
pseudorapidity acceptance of $2<\eta<5$ for charged particles.  The
detector includes a high precision tracking system, which consists of a
silicon vertex detector and dedicated tracking planes. The tracking planes consist of silicon microstrip detectors in the region with high charged-particle flux close to the beam pipe and straw tube detectors which provide coverage up to the edge of the LHCb geometrical acceptance. The tracking planes are located either side of the dipole magnet to allow the measurement of the momenta of charged particles as they traverse the detector.
Excellent particle identification capabilities are provided by two ring imaging Cherenkov detectors which allow charged pions, kaons and
protons to be distinguished from each other in the momentum range 2--100 \gevc. The energy of particles traversing the detector is measured using a calorimeter system which is sensitive to photons and electrons, as well as hadrons. Muons are identified using a dedicated detector system.

The experiment employs a multi-level trigger comprised of a hardware trigger which uses
information from the calorimeter and muon system and a software trigger which performs a full reconstruction of the event, including tracks and vertices.

The simulated events used in this analysis are produced using the \pythia 6.4 generator \cite{Sjostrand:2006za}, with a choice of parameters specifically configured for \lhcb \cite{LHCb-PROC-2010-056}. The \evtgen package \cite{Lange:2001uf} describes the decay of $b$-hadrons and the \geant toolkit \cite{Agostinelli:2002hh} simulates the detector response, implemented as described in Ref. \cite{LHCb-PROC-2011-006}. QED radiative corrections to the \BsToKK decay are generated with the \photos package \cite{Golonka:2005pn}.

section{Trigger and event selection}
\label{sec:Data}
At \lhcb, $b$-hadrons are produced with an average momentum of around 100 \gevc and have decay vertices displaced from the primary interaction
vertex. Combinatorial background candidates, produced by the random combination of tracks, tend to have low momentum and originate from a primary \textit{pp} collision vertex. These features are typically exploited to select $b$-hadrons and reject background. The distance of closest approach (impact parameter) of $b$-hadron decay products to any primary vertex is a particularly important discriminant in the trigger because it is an order of magnitude faster to compute than the momenta of the same decay products. For this reason, the majority of triggers for hadronic $b$-hadron decays begin by selecting tracks with a significant displacement from any primary vertex. However, such requirements introduce a time-dependent acceptance which biases the decay time distribution of the selected $b$-hadron candidates and a significant investment of effort is often required to correct for this bias.

The analysis presented here uses an approach that selects $b$-hadrons without biasing the decay time distribution, other than trivially through a simple minimum decay time requirement, limiting the systematic uncertainties associated with correcting for any time-dependent acceptance effects. This is achieved using neural networks based on the NeuroBayes package \cite{Feindt:2006pm} in the software trigger and event selection. Neural networks have advantages over traditional ``cut-based'' approaches since they are able to exploit the correlations between variables in order to increase signal purity, allowing $b$-hadrons to be selected without resorting to requirements on impact parameters or flight distance.

The \lhcb software trigger has two stages which run sequentially. Due to restrictions on processing time it is not possible to employ a neural network in the first level of the software trigger. Instead, only tracks that are not used in the first level decision are passed to the second trigger level in order to avoid a potential bias. These tracks are required to pass a loose pre-selection with requirements on their momenta, transverse momenta and track fit quality. The tracks are then combined to form \B meson candidates, using a kaon mass hypothesis for both tracks, and further requirements are made on the distance of closest approach of the two tracks to each other, the mass of the resulting candidate, the helicity angle of the tracks in the \B meson rest frame and the quality of the decay vertex fit.

After this pre-selection the candidates pass through a first neural network, trained on simulated \BsToKK, \BdToKpi and background events, which uses the momenta and transverse momenta of the tracks and \B meson candidate, the distance of closest approach of the two tracks, helicity angle, the $\chi^2$ of the vertex fit and the uncertainty on the fitted \B meson mass to discriminate between signal and background. After this stage the data rate is reduced to a level such that each event may be fully reconstructed, including information from the particle identification system. A second network, trained on the same simulated events, uses the information presented to the first network along with particle identification information to further increase the purity of \B mesons in the selected candidates.

Roughly half way through 2011 the luminosity delivered by the LHC accelerator increased to a level such that it was necessary to require that the decay time of \B meson candidates exceeded 0.3 ps in order to keep the trigger rate within acceptable limits. This requirement only biases the decay time distribution in a trivial way, except through a possible difference in the decay time resolutions of the trigger and offline reconstruction software. 

After the trigger, the tracks associated to the selected candidates are removed from the primary vertex fit to avoid a potential bias in the measured decay time. The purity of signal candidates is then further enhanced using two additional sequential neural networks. The first network is trained using simulated events and combines the same information used by the trigger networks along with particle identification information, the energy of each track from the calorimeter, the probability that either track is formed from the association of random hits in the detector and the $\chi^2$ per degree of freedom for both track fits. This network benefits from the more detailed full event reconstruction which is not available in the trigger. 

The second network is trained on the data recorded in 2011 using \textit{sWeights} \cite{sWT}, which are calculated in a window around the signal peak and in the upper sideband region ($5.45 < m_{\Kp\Km} < 5.85 \gevcc$) of the invariant mass spectrum. The \textit{sWeights} are obtained from a fit to the invariant mass spectrum of the candidates and the neural network uses them to discriminate between signal and background. This network uses the output of the first network as input, all the input variables used by the first network, the uncertainty on the decay time of the \B meson candidate and the impact parameter of the \B meson candidate with respect to the primary interaction vertex. Only candidates with a decay time of $\tau > 0.3$ ps are used in the network training.

The event selection is determined by making a requirement on the output of this second neural network that maximises the metric $s/\sqrt{s + b}$, where $s$ is the number of signal decays in the  region $ 5.05 < m_{\Kp\Km} < 5.85 \gevcc$ and $b$ is the number of background combinations.

The trigger and offline software reconstruct \B meson decay times with different resolutions. Potential ``edge-effects'' introduced by the trigger requirement that $\tau > 0.3$~ps are avoided by requiring that candidates satisfy $\tau > 0.5$ \ps in the final event selection. The contribution from the \BdToKpi and \BsToKK modes are separated by demanding tight requirements on the particle identification properties of the final state particles. A small level of contamination from decays of $\Lambda_b$ baryons is further suppressed by demanding that the final state particles are not compatible with the proton hypothesis. 

\section{Analysis of the effective $\boldsymbol{ B_s^0 \rightarrow K^+ K^- }$ lifetime}
\label{sec:Ana}
The effective \BsToKK lifetime is evaluated using an unbinned log-likelihood fit. A fit to the invariant mass spectrum is performed to determine the \textit{sWeights} that are used to isolate the \BsToKK decay time distribution from the residual background. The \BsToKK signal component is described by a Gaussian function. The background contamination from partially reconstructed \B meson decays is described by a further Gaussian function and the combinatorial background is described by a Chebychev polynomial with one free parameter. It should be noted that the kaon mass is assigned to both final state particles in the vertex fit and hence the reconstructed \BdToKpi mass is shifted towards higher values than the nominal mass, creating an asymmetric distribution. The \BdToKpi signal component is therefore described by a Crystal Ball function \cite{Skwarnicki:1986xj} with the tail on high mass side. The parameters of this distribution are fixed using a fit to the independent \BdToKpi sample, separated using particle identification information.

\begin{figure}[h]
\label{fig:BsKKMassTau}
\begin{center}
 \includegraphics[width=0.49\textwidth]{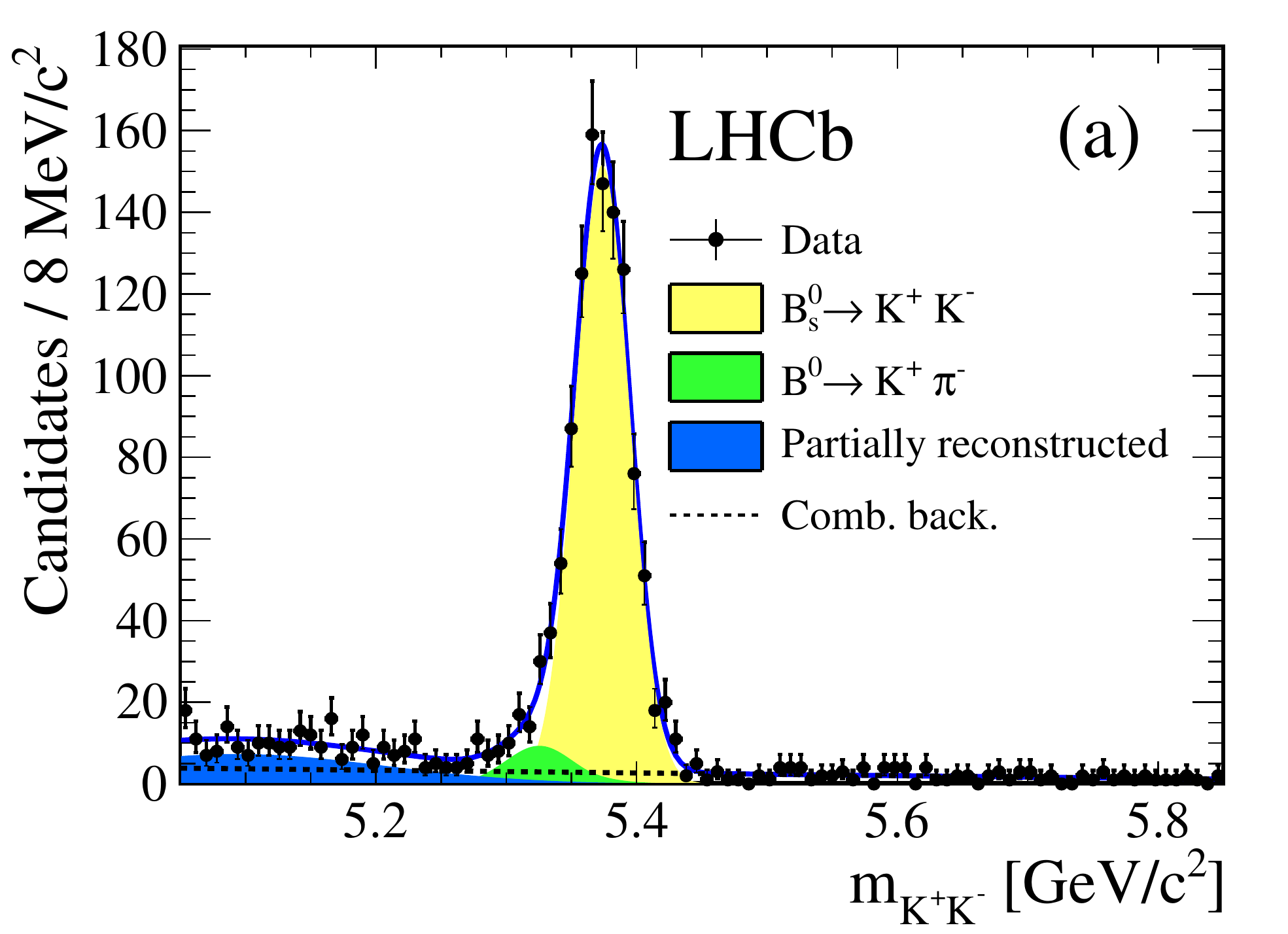} 
 \includegraphics[width=0.49\textwidth]{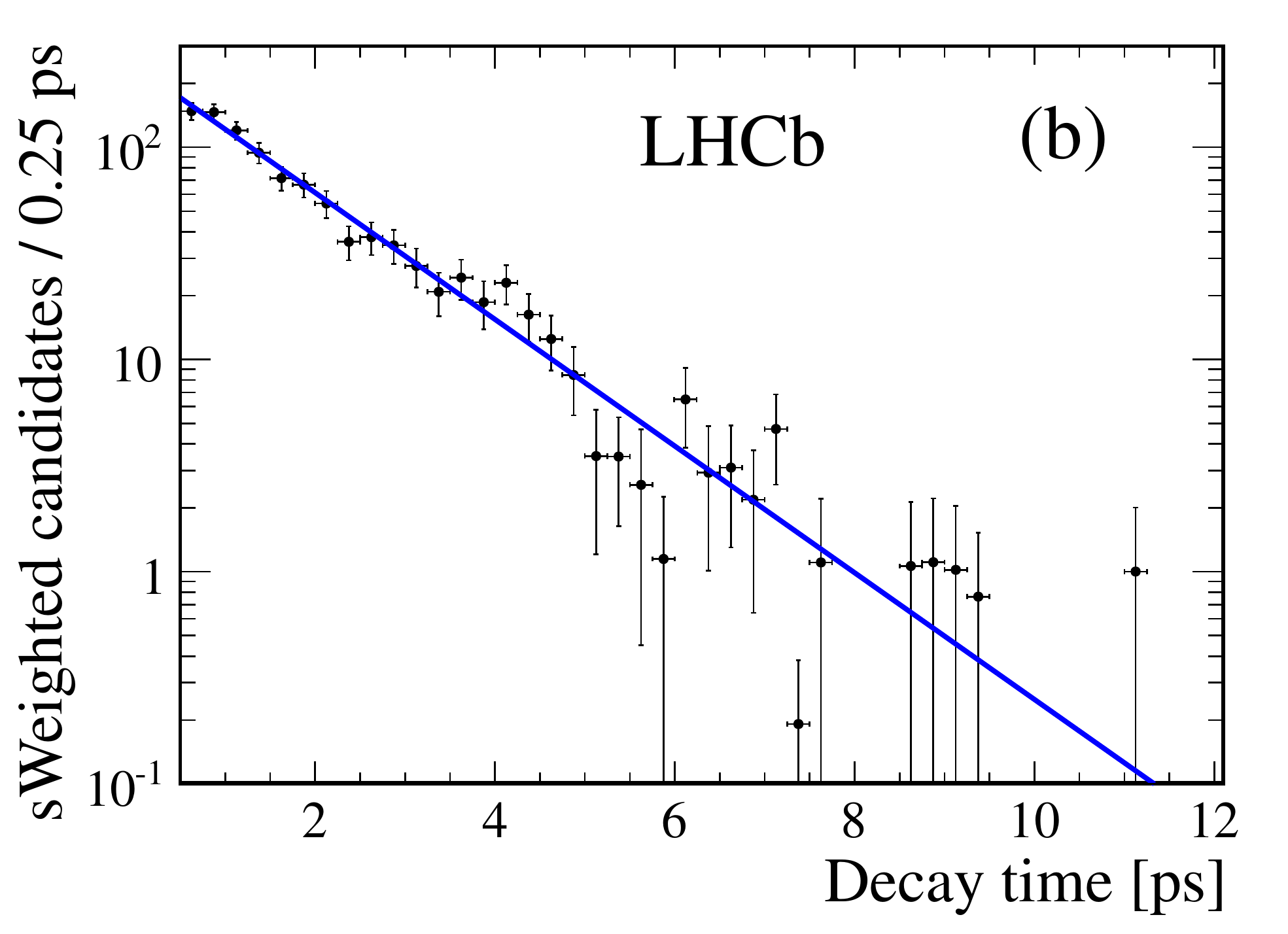}
\caption{ (a) Invariant mass spectrum for all selected \BsToKK candidates.
(b) Decay time distribution of \BsToKK signal extracted using \textit{sWeights} and the fitted exponential function.}
\end{center}
\end{figure}

The fit finds $997 \pm 34$ \BsToKK decays and $78 \pm 17$ \BdToKpi decays in the data with $253 \pm 25$ and $169 \pm 20$
combinatorial background and partially reconstructed combinations respectively. Figure \ref{fig:BsKKMassTau}(a) shows the resulting
invariant mass spectrum for \BsToKK candidates.

Using the \textit{sWeights} returned by the mass fit, the \BsToKK decay time distribution is extracted from data using the \textit{sPlot} technique \cite{sWT}. Since there is no acceptance bias to correct for, the lifetime is determined using a fit of the convolution of an exponential and Gaussian function to account for the resolution of the detector. The mean of the Gaussian function is fixed to zero and its width is fixed to the expected resolution from simulated events, which is $\sigma_{t} = 0.04$~ps.

The effective \BsToKK lifetime is found to be
\begin{displaymath}
\tauBsToKK = 1.455 \pm 0.046 \; \mathrm{(stat.)} \; \ps.
\end{displaymath}
Figure \ref{fig:BsKKMassTau}(b) shows the corresponding fit to the decay time distribution of \BsToKK signal.

Since the decay \BdToKpi has similar kinematics, it can be used
as a control mode. However, since the kaon mass hypothesis is assigned to both tracks, the measured decay time is biased to larger values for \BdToKpi. To avoid this bias a fit is made to the reduced decay time, which is defined as the decay time divided by the invariant mass. This quantity is independent of the mass assigned to the two tracks and is also unbiased by the selection, following an exponential distribution with decay constant equal to $m_{\Bz}/\tauBd$.

Using the value of the \Bz mass \cite{Nakamura:2010zzi}  as input, the \Bz lifetime is found to be
\begin{displaymath}
\tauBd = 1.536 \pm 0.031 \; \mathrm{(stat.)} \; \ps
\end{displaymath}
which agrees with the current world-average
$\tauBd = 1.519 \pm 0.007$ ps \cite{Nakamura:2010zzi}.

\FloatBarrier

\section{Evaluation of systematic uncertainties}
\label{sec:Syst}

A wide range of effects that can influence the measurement of the effective \BsToKK lifetime has been evaluated. The individual 
contributions to the systematic uncertainties are described below and their estimated values are summarised in Table \ref{table:ResultBsKK}.

\begin{figure}[h]
\begin{center}
\includegraphics[width=0.49\textwidth]{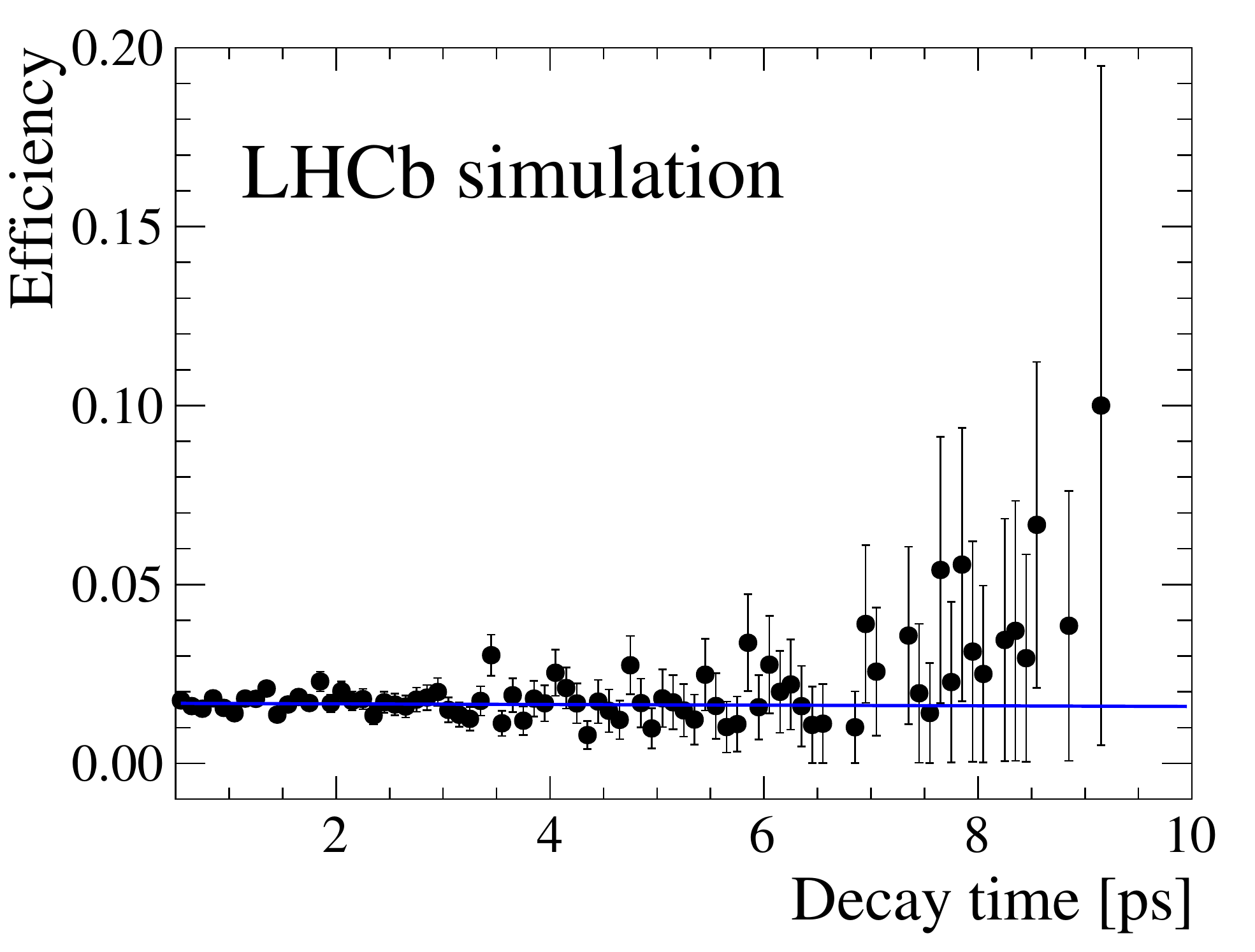}
\caption{Combined efficiency of \lhcb trigger, selection neural networks and particle identification requirements as a function of decay time for simulated \BsToKK signal candidates.}
\label{fig:TauRatioOffline}
\end{center}
\end{figure}

The key principle of this analysis is that the trigger and event selection do not bias the decay time distribution of the selected \BsToKK candidates other than in a trivial way through a minimum decay time requirement. This has been tested extensively using simulated events at each stage of the selection process to demonstrate that no step introduces a time-dependent acceptance. Figure \ref{fig:TauRatioOffline} shows the efficiency of the full trigger and event selection as a function of decay time for simulated \BsToKK candidates. The graph is fitted with a first order polynomial with a gradient of $-0.09 \pm 0.30$~ns$^{-1}$ consistent with a uniform acceptance. Possible discrepancies between simulated and real events are considered by comparing the distributions of variables used by the neural networks and good agreement is observed. The available quantity of simulated events limits any non-zero gradient in the acceptance to within 0.30~ns$^{-1}$. This limit is used to evaluate the shift in the measured effective lifetime due to the presence of a linear acceptance and a negligible deviation is observed and is not considered any further.

Studies using simulated events have demonstrated that the efficiency with which tracks are reconstructed decreases as the impact parameter of the track with respect to the beam line (IP$_z$) increases. This introduces a decay time acceptance that may bias the measured lifetime. Such a systematic bias has been evaluated using a combination of data and simulated events. First, the effective lifetime of simulated \BsToKK signal candidates is found after reconstruction to deviate by 5~fs from the generated value. Second, the tracking efficiency is parametrised as a function of IP$_z$ using simulated events. The calculated efficiency is then applied as a weight to events in data according to their IP$_z$ values and the effective lifetime is evaluated. This produces a deviation of 4 fs with respect to the unweighted events. The larger of these two shifts is taken as the systematic uncertainty introduced by the reconstruction acceptance.

The invariant mass distribution of \BsToKK signal candidates is modelled using a Gaussian function. Potential systematic effects due to this parametrisation are evaluated by using the sum of two Gaussian functions to model additional resolution effects and separately a Crystal Ball function \cite{Skwarnicki:1986xj} to model final state radiation. Additionally the background parametrisation is checked by replacing the first order Chebychev polynomial with an exponential function. All these changes shift the measured lifetime by approximately 1 fs which is taken as the systematic uncertainty.

The decay time distribution is fitted with an exponential function convolved with a Gaussian function to model detector resolution, where the resolution is fixed to the value obtained from simulated events. As a cross-check, the fit is performed with the resolution parameter allowed to vary and also using a simple exponential function without attempting to model detector resolution. No deviation from the default measurement of the effective lifetime is observed in either case.

The effective \BsToKK lifetime measurement has been evaluated using an alternative method which makes a simultaneous fit to the invariant mass and decay time distributions. This approach requires a parametrisation of the background decay time distribution since the \textit{sPlot} technique is not used. Both methods give equivalent numerical results. 

A wide range of different approaches to the training of the neural network have been tested, as well as the influence of different alignment and calibration settings and the number of simultaneous primary interactions in the detector. All results obtained in these checks are consistent with the result of the default analysis.

The measured decay times of \B meson candidates are determined from the distance between the primary interaction and the secondary decay vertex in the silicon vertex detector. A systematic bias may therefore be introduced due to uncertainty on the \lhcb length scale. This effect is estimated by considering the uncertainty on the length scale from the mechanical survey, thermal expansion and the current alignment precision. The uncertainty on the length of the detector along the beam-line is determined to be the dominant effect and a corresponding systematic uncertainty is assigned.

The effective lifetime is obtained by fitting a single exponential function to the distribution given by Eq.~\ref{eqn:dblExp}. However, the requirement that the decay time be greater than 0.5~ps diminishes the \GL component relative to the \GH component in the decay time distribution. This effect has been evaluated using simulated events and a deviation of 1 fs from the result of a fit to the full decay time range is observed.

If the production rates, $R$, of \Bs and \Bsb mesons are not equal then an additional oscillatory term is introduced into the decay time distribution given in Eq.~\ref{eqn:dblExp}, proportional to the production asymmetry $A_P \equiv \left[R(\Bsb) - R(\Bs)\right]/\left[R(\Bsb) + R(\Bs)\right]$. This term may alter the measured effective lifetime. Since the \Bs meson shares no valence quarks with the proton $A_P(\Bs)$ at \lhcb is expected to be small. Making the conservative assumption that the $|A_P(\Bs)| = |A_P(\Bz)| = 0.01$ \cite{PhysRevLett.108.201601} we find a shift from the expected value of the effective lifetime of 2 fs using simulated events. This value is assigned as the systematic uncertainty.

\begin{table}[h] 
\label{table:ResultBsKK} 
\caption[Syst. Error] {Contributions to the systematic uncertainty on the effective \BsToKK lifetime measurement. The total uncertainty is calculated by adding the individual contributions in quadrature.} 
\begin{center} 
\begin{tabular}{ |l|c| } \hline 
Systematic sources &  Uncertainty on \tauBsToKK [fs] \\ \hline 
Reconstruction efficiency   & $5$ \\
Signal model &   $1$\\ 
Background model &  $1$\\ 
Length scale   & $1$ \\
Minimum decay time requirement & $1$ \\ 
Production asymmetry           & $2$ \\ \hline
Total & $6$   \\ \hline 
\end{tabular} 
\end{center} 
\end{table}


\section{Conclusions}
Two-body charmless \B decays offer a rich phenomenology to explore the phase structure of the CKM matrix and to search for manifestations of physics beyond the SM.  The effective lifetime of the decay \BsToKK is of considerable theoretical interest as it is sensitive to new particles entering at loop level. A measurement of this quantity is made possible by the excellent particle identification capabilities of the \lhcb experiment.

The effective lifetime of the decay mode \BsToKK is measured using 1.0 \invfb of data recorded by the \lhcb detector in 2011. A key element of this analysis is that the trigger and event selection selects \B mesons without biasing the decay time distribution. This is achieved using a series of neural networks. Although this dedicated trigger has a lower efficiency compared to the one used in the previous \lhcb measurement \cite{TauBhhLHCbPLB}, it has the advantage of avoiding systematic uncertainties related to the depletion of candidates at low decay times and provides an independent approach to measuring the \BsToKK effective lifetime. It is measured as
\begin{displaymath}
\tauBsToKK = 1.455 \pm 0.046 \; \mathrm{(stat.)} \pm 0.006 \;  \mathrm{(syst.) \; ps,}
\end{displaymath}
in good agreement with the SM prediction of $1.40 \pm 0.02$ ps and with the measurement on data recorded by LHCb in 2010 of $1.440 \pm 0.096 \; \mathrm{ (stat.)} \pm 0.008 \; \mathrm{ (syst.)} \pm 0.003 \; \mathrm{ (mod.)}$~ps~\cite{TauBhhLHCbPLB}.

\section*{Acknowledgements}

\noindent We express our gratitude to our colleagues in the CERN accelerator
departments for the excellent performance of the LHC. We thank the
technical and administrative staff at CERN and at the LHCb institutes,
and acknowledge support from the National Agencies: CAPES, CNPq,
FAPERJ and FINEP (Brazil); CERN; NSFC (China); CNRS/IN2P3 (France);
BMBF, DFG, HGF and MPG (Germany); SFI (Ireland); INFN (Italy); FOM and
NWO (The Netherlands); SCSR (Poland); ANCS (Romania); MinES of Russia and
Rosatom (Russia); MICINN, XuntaGal and GENCAT (Spain); SNSF and SER
(Switzerland); NAS Ukraine (Ukraine); STFC (United Kingdom); NSF
(USA). We also acknowledge the support received from the ERC under FP7
and the Region Auvergne.

\newpage

\bibliographystyle{LHCb}
\bibliography{main}

\end{document}